# Selective temperature sensing in nanodiamonds using dressed states

N. M. Beaver, P. Stevenson[*]

Department of Physics, Northeastern University, Boston, MA

*p.stevenson@northeastern.edu

**Abstract:**

Temperature sensing at the nanoscale is a significant experimental challenge. Here, we report an approach using dressed states to make a leading quantum sensor – the nitrogen vacancy (NV) center in diamond – selectively sensitive to temperature, even in the presence of normally-confounding magnetic fields. Using an experimentally straightforward approach, we are able to suppress the magnetic sensitivity of the NV center by a factor of seven while retaining full temperature sensitivity and narrowing the NV center linewidth. These results demonstrate the power of engineering the sensor Hamiltonian using external control fields to enable sensing with improved specificity to target signals.

**Introduction**

Measuring temperature at the nanoscale is a vital experimental capability in both fundamental contexts, such as understanding microscopic thermodynamics [1,2], and application-focused uses, such as device testing [3,4]. Several experimental approaches have been developed to meet this challenge [2,5,6], but often encounter the problem of multiple overlapping signals (e.g., temperature and current [3], temperature and strain [2]) which makes assigning the temperature-specific response challenging. Here, we show that judicious application of external control fields to a quantum sensor can be used to improve the selectivity of our sensor to temperature signals.

The diamond nitrogen vacancy (NV) center is a point defect in diamond, which provides an atom-size sensor capable of detecting magnetic fields, electric fields, and temperature. The NV center is fluorescent, paramagnetic, and has an optical cycle which allows both optical polarization of the system into a single spin level and optical readout of the spin state. Variations in the spin state, either from magnetic fields (Zeeman shift) or from temperature (temperature-dependent variations in the zero-field splitting [7–12]) can be determined by probing the transition energies of the spin states.

NV centers are excellent magnetometers [13–20] with sensitivities to DC fields reaching the picotesla level [21]; in the context of temperature sensing, however, this sensitivity to magnetic fields presents a challenge since even small changes in the background magnetic field can obscure the signal from temperature changes. For example, even in simple electrical devices, temperature changes on the order of 10°C (<1MHz frequency shift) can be obscured by magnetic signals on the order of 2mT (>50MHz frequency shift) [3].

Several strategies have been explored to improve the selectivity of the NV center response to temperature, such as the thermal echo [22] and D-Ramsey pulse sequences [23], spectral hole burning [24], and simultaneous measurement of magnetic field and temperature [3]. An alternative to these approaches utilizes "dressed states" [25,26], where externally applied, time-varying signals can be used to tailor the Hamiltonian of the system, making it *selectively* sensitive to different signals. Here, we demonstrate the generation of dressed states in nanodiamonds through the application of a single additional radiofrequency (RF) drive in the megahertz range. Our goal is to use this capability to make the NV center selectively sensitive to temperature by suppressing the magnetic field response and explore whether this can improve the intrinsic sensing performance.

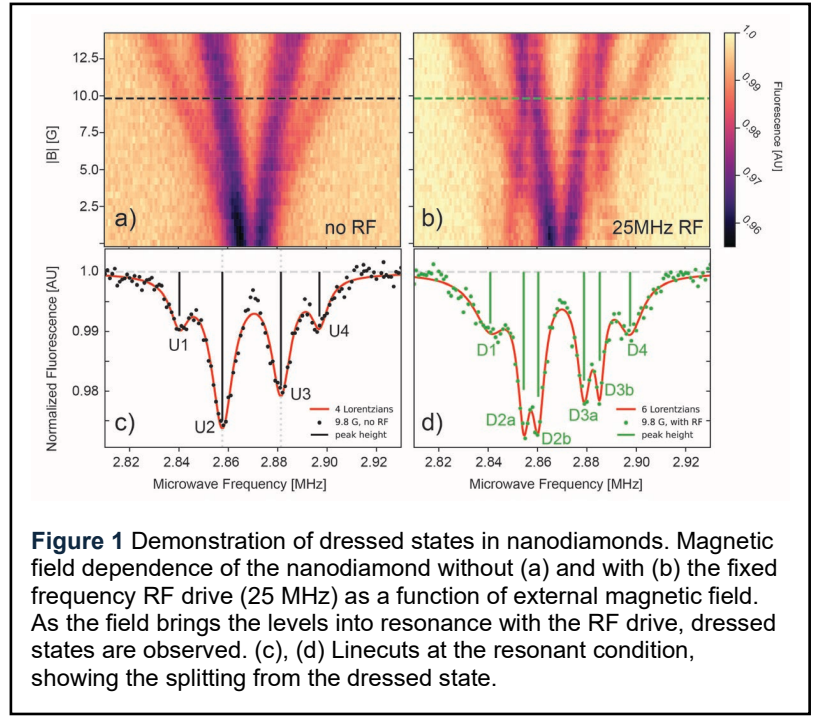

**Figure 1** Demonstration of dressed states in nanodiamonds. Magnetic field dependence of the nanodiamond without (a) and with (b) the fixed frequency RF drive (25 MHz) as a function of external magnetic field. As the field brings the levels into resonance with the RF drive, dressed states are observed. (c), (d) Linecuts at the resonant condition, showing the splitting from the dressed state.

While dressed states have been explored extensively in NV centers in large single crystal substrates, here, we focus on nanodiamonds. Nanodiamonds can easily be integrated into a variety of systems ranging from living cells [27,28] to transistors [3], but they present unique challenges in terms of paramagnetic background [29], strain, and surface noise sources [30]. We extend previous descriptions of dressed states in large single crystal samples [25,29,31–40,40,41] to these inexpensive, easy-to-integrate nanodiamonds, demonstrating their utility as nanoscale temperature sensors and developing further routes for optimization.

**Dressed states with nanodiamonds**

The ground-state Hamiltonian for an NV center with applied magnetic field ($\vec{B}$), including the effect of off-axis strain is

$$\hat{H} = D_{GS}\hat{S}_z^2 + E_x(\hat{S}_x^2 - \hat{S}_y^2) + \sum_{i=x,y,z} \gamma_e \hat{S}_i B_i \qquad (1)$$

where $S_x, S_y, S_z$ are the spin-1 operators; $D_{GS}$ is the zero-field splitting; $E_x$ is the off-axis strain; $\gamma_e = 2\pi\ 2.8\ \text{MHz/G}$ is the gyromagnetic ratio for the NV center; and $B_x, B_y, B_z$ are the components of the static bias field; $z$ is defined to lie along the NV quantization axis (the N-V direction).

We can alter this Hamiltonian by applying a time-dependent perturbation to the system. We apply a MHz radiofrequency drive to "dress" the states [25,29,31,33,34,38–44] and apply a GHz microwave drive to directly drive transitions between states. Under these conditions the Hamiltonian becomes:

$$\hat{H} = D_{GS}\,\hat{S}_z^2 + E_x(\hat{S}_x^2 - \hat{S}_y^2) + \sum_{i=x,y,z} \gamma_e(B_i + B_{i_{MW}}\cos(\omega_{MW}t) + B_{i_{RF}}\cos(\omega_{RF}t))\hat{S}_i \qquad (2)$$

where $B_{i_{RF(MW)}}$ is the $i$th component (x, y, z) of the MHz (GHz) driving field; $\omega_{RF(MW)}$ is the angular frequency of the RF (MW) drive; and $t$ is time. The solutions of this Hamiltonian under different drive powers and frequency regimes have been discussed extensively [25,29,31,34–40,42–44], but generally show a modification of the observed spectrum under the effect of the RF drive field which results in the appearance of additional peaks at frequencies determined by the RF drive.

First, we generate and characterize dressed states in NV center-containing nanodiamonds. Figure 1(a-b) shows the evolution of the NV center spectrum with and without a dressing field at a fixed RF frequency (25 MHz) as an external bias field is applied to a nanodiamond. Without the applied RF field (Figure 1a), we observe the typical Zeeman shift: there are four distinct orientations of the NV center, however, our applied magnetic field is approximately aligned with one of the orientations, resulting in the other three experiencing approximately the same magnetic field. As the field increases, the transitions split further apart. The splitting at no applied field arises from the transverse strain term in the Hamiltonian ($E_x$).

However, when a dressing field is applied (Figure 1b), we see a different behavior in the inner sets of peaks when the peak separation is equal to the frequency of the dressing field. Specifically, when the separation of the two inner peaks matches the separation of the RF drive (25 MHz), we see each peak split into two sub-peaks (Figure 1d) with an avoided crossing-like structure. We observe that the linewidth of the peaks decreases in this dressed state (9.2 MHz to 6.4 MHz, see Table 1); this has previously been observed in bulk diamond samples, where the line narrowing is greater, improving linewidth from >6 MHz to <1 MHz [25], though this effect had contributions from the orientation of the magnetic field as well as the dressed states [45]. We discuss potential origins and limitations of this reduced narrowing later. We are able to generate these dressed states in many nanodiamonds under a range of different magnetic field conditions (Tables 1, 2), highlighting their broad applicability.

Figure 2 shows the effect of varying drive frequency and power at fixed magnetic fields. While the frequency-swept spectrum (Figure 2a) is congested because of the detailed structure of the dressed state and multiple NV orientations, we are still able to describe the observed peak positions using the model described in Ref. [40]

We also consider the effect of RF (dressing) power on the linewidth of the peaks. We observe the expected increase in peak splitting as the Rabi frequency of the RF field increases (Figure 2b), and at higher powers

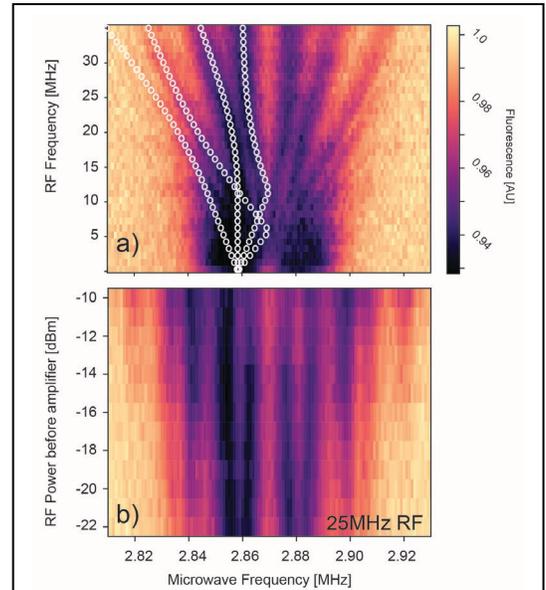

**Figure 2** (a) RF frequency dependence of the dressed states. Overlaid are the simulated one-photon transitions calculated from Ref 40. Transitions on the high frequency side are omitted for clarity. (b) RF-power dependence of ODMR at the avoided crossing. We observe an increase in the number of peaks (multi-photon transitions) as the power increases.

observe the formation of higher-order sidebands [31,32,35,40,46], confirming the dressed state nature of the features. Here, we see that the contrast of the central peaks saturates quickly at lower RF powers, but the linewidth of the dressed state peaks does not increase with increasing RF power. This is in contrast to conventional power-broadening effects in two-level systems, where an increase in Rabi frequency broadens the observed transition. Practically, this simplifies the experimental implementation of this approach, where careful optimization of the dressing power is less critical than other parameters [47]. We contrast this with other dressed state generation schemes which drive multiple transitions with microwave frequencies [26]; in these schemes, the transition can be broadened at high drive power, reducing sensitivity.

| Nanodiamond # | Label | Bias field [G] | $\eta_T$ [K/$\sqrt{\text{Hz}}$] | linewidth [MHz] | Contrast [%] | $\partial D/\partial T$ [kHz/K] |
|---|---|---|---|---|---|---|
| 1 | No RF drive | 7.0 | 3.2 | 9.2 | 2.5 | -92.7 ± 2.3 |
| 1 | With RF drive (22 MHz) | 7.0 | 2.9 | 6.4 | 2.0 | -93.3 ± 5.3 |

Table 1: Fit parameters and calculated temperature sensitivities for nanodiamond #1 with and without the dressed state drive.

**Temperature sensing with dressed states**

Next, we consider the prospects of using dressed states for temperature sensing. NV centers have been used to sense temperature in a wide range of contexts through a broad range of mechanisms [27,48–54], where the high spatial resolution, high thermal conductivity, and excellent chemical resistance make diamond an attractive host material for these sensors. For ODMR-based sensing, the temperature sensitivity is defined as: [26,55]

$$\eta_T = \frac{1}{\frac{\partial D_{GS}}{\partial T}} \frac{\Delta\omega}{C\sqrt{R}} \tag{3}$$

where $\eta_T$ is the temperature sensitivity in K/$\sqrt{\text{Hz}}$, $\Delta\omega$ is the linewidth in Hz, $\partial D_{GS}/\partial T$ is the temperature shift in Hz/K, $C$ is the ODMR peak contrast in normalized counts, and $R$ is the count rate in Hz. As outlined in Table 1, we find a modest improvement in the calculated sensitivity under the dressed state drive. The decrease in linewidth improves the sensitivity, but the concurrent decrease in the ODMR *amplitude* (contrast) for each dressed peak means we expect only a modest improvement from 3.6 K/$\sqrt{\text{Hz}}$ (bare) to 2.9 K/$\sqrt{\text{Hz}}$ (dressed). We note that the emission count rate in our experiments, $R$, saturates our detector so we attenuate the fluorescence with filters. We use the attenuated count rate for determining these sensitivities, meaning these numbers are an underestimate of the true sensitivities possible with our system.

Figure 3 shows ODMR data taken at 25°C and 50°C for the bare and dressed states, where we can directly see the shift in peak frequency with temperature. This shift to lower frequency with increasing temperature arises from changes in the strain in the diamond lattice, which in turn changes the zero-field splitting parameter $D_{GS}$. We compare the response of the nanodiamonds to temperature under bare and dressed conditions, finding the thermal response to be the same,

≈ 93 kHz/K . We note that this coefficient is higher than the standard value of 75 kHz/K reported for bulk diamonds [7], but comparable to previous reports in nanodiamonds [56,57]. As a point of comparison, we note that a 10K temperature change would result in a frequency shift equal in magnitude to ~3μT, which is comparable to local variations in the earth's magnetic field, which motivates our efforts to suppress the magnetic field sensitivity.

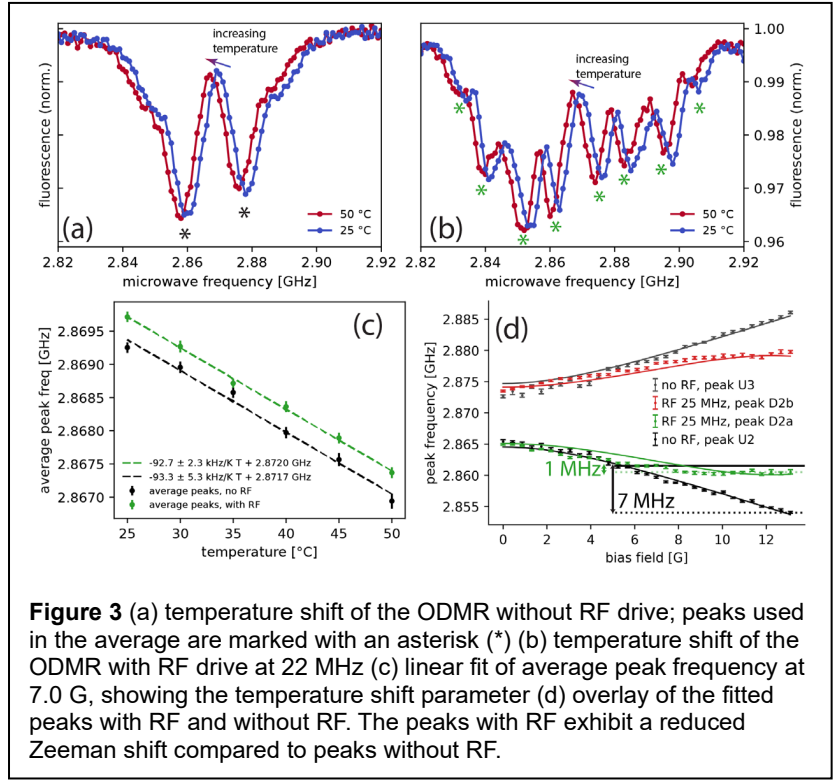

**Figure 3** (a) temperature shift of the ODMR without RF drive; peaks used in the average are marked with an asterisk (*) (b) temperature shift of the ODMR with RF drive at 22 MHz (c) linear fit of average peak frequency at 7.0 G, showing the temperature shift parameter (d) overlay of the fitted peaks with RF and without RF. The peaks with RF exhibit a reduced Zeeman shift compared to peaks without RF.

The magnetic field dependence of the bare and dressed states show distinctly different behaviors. The bare NV centers show the typical linear Zeeman shift of 2.8 MHz/G (some deviation from linear behavior is observed at low fields where transverse strain terms dominate the Hamiltonian). The dressed NV centers, however, display a different behavior; over the range 6 to 14 G, we see a shift of only ~1MHz. This frequency shift is a factor of seven lower than the bare NV centers, while still retaining the full temperature sensitivity. This result is consistent with previous reports where dressed states have been used to increase the coherence time of qubits by minimizing their sensitivity to magnetic noise [41,58–63]; here, we envision using this insensitivity to improve our ability to *selectively* sense temperature changes even in the presence of magnetic field variations.

To model the peak frequency for the dressed states (Figure 3d), we use the following equation for the four primary peaks of the lowest-order dressed state transitions [40]:

$$\omega_{MW} = D_{GS} \pm \frac{1}{2}\omega_{RF} \pm \sqrt{\left(\sqrt{(\gamma_e B \cos\theta)^2 + E_x^2} - \frac{1}{2}\omega_{RF}\right)^2 + \left(\frac{\Omega_{RF}}{2}\right)^2} \quad (4)$$

where $\omega_{RF(MW)}$ is the angular frequency of the RF (MW) drive, $D_{GS}$ is the zero-field splitting; $E_x$ is the off-axis strain; $\gamma_e = 2\pi \times 2.8$ MHz/G is the gyromagnetic ratio for the NV center, $B\cos\theta = B_z$ is the on-axis component of the bias field, and $\Omega_{RF} = \gamma_e B_{z_{RF}}$ is the Rabi frequency for the RF drive. The model for the dressed states we plot in Figure 3d has no free parameters: all values were fixed from other measurements (summarized in the SI). Specifically, the bare-state ODMR allows us to determine the relative orientation ($\cos\theta = -\frac{1}{3}$) and zero-field parameters ($D_{GS}$ and $E_x$), while the Rabi frequency is determined from the dressed-state splitting ($\Omega_{RF}$, Figure 1d). The suppression of magnetic field sensitivity works over a limited range; large changes in field shift the NV transitions out of resonance with the dressing field.

Under our experimental conditions, however, we find this is able to suppress the sensitivity to magnetic field by a factor of 7 (Figure 3d), which could be further improved with larger RF drive power (larger $\Omega_{RF}^2$).

| Nanodiamond # | Bias field [G] | best $\eta_T$ [K/√Hz], no RF | best $\eta_T$ [K/√Hz], RF |
|---|---|---|---|
| 1 | <1 G | 2.7 | 1.8 |
| 2 | <1 G | 4.1 | 2.3 |
| 3 | <1 G | 4.4 | 2.7 |
| 4 | <1 G | 5.5 | 2.8 |

Table 2: temperature sensitivity for other nanodiamonds without applied bias field.

**Simulations and optimizing material parameters for dressed state sensing**

While the low cost and ease of integration of nanodiamonds makes them appealing sensors, the sensitivities we report here (and in prior works [49,56,64,65]) fall behind state-of-the-art large single crystal diamond sensors [49,64,66,67]. Additionally, while we see a linewidth narrowing (Figure 1) in the dressed state, this change in linewidth is less pronounced than previous descriptions in single crystal substrates. To explore the origins of these effects, we numerically simulate spectra using the Hamiltonian in Eq. 2 with a Monte Carlo approach to determine how disorder in different degrees of freedom impacts the bare and dressed state of our system.

Figures 4a and 4b show representative spectra for the bare and dressed state, where, as expected, the presence of the RF drive causes multiple peaks to appear. We introduce disorder in different terms of the Hamiltonian corresponding to different physical parameters and determine how the linewidth of the resulting ensemble changes. We characterize how the system changes with longitudinal and transverse strain disorder ($D_{GS}$, $E_x$, respectively) and on-axis and off-axis magnetic field disorder ($B_z$, $B_x$ respectively).

| Parameter | Value (units) | Description |
|---|---|---|
| $N$ | 4 | Floquet order |
| $D_{GS}$ | $2\pi$ 2.87 GHz | Zero-field splitting |
| $E_x$ | $2\pi$ 5.0 MHz | Off-axis strain |
| $\Omega_{x_{MW}}$ | $2\pi$ 0.24 MHz | Microwave (GHz) Rabi frequency in x-direction ($\gamma_e B_{x_{MW}}$) |
| $\Omega_{y_{MW}}$ | $2\pi$ 0.24 MHz | Microwave (GHz) Rabi frequency in the y-direction ($\gamma_e B_{x_{MW}}$) |
| $\omega_{RF}$ | $2\pi$ 10.0 MHz | RF (MHz) drive angular frequency |
| $\Omega_{RF}$ | $2\pi$ 4.0 MHz | RF (MHz) Rabi frequency in z-direction ($\gamma_e B_{z_{RF}}$) |

Table 3: parameters used for Floquet method cw-ODMR in Monte Carlo simulations.

From these simulations, we observe that variations in $E_x$, $B_z$, and $B_x$ give rise to line-broadening which is effectively suppressed with the dressed state (Figure 4), in line with previous observations [25,37,42]. $B_x$ has only a small effect on transition frequency in the regime we consider [68,69], so we neglect further discussion of this term. Variations in both $E_x$ (from strain inhomogeneity [70,71]) and $B_z$ (from paramagnetic impurities [29,72–74]) are both expected to be present in our samples; however, our simulations suggest we should expect more significant linewidth narrowing in our experimental data.

To explain this discrepancy, we consider two alternative sources of broadening. First, inhomogeneity in the $D_{GS}$ parameter ("$D$-strain") would result in broadening which is *not* addressed by the dressed state (Figure 4f). Variations in this parameter on the order of 3 MHz have been reported across different nanodiamonds [75], suggesting this could explain, in part, some of the effects we see.

Second, we consider the presence of multiple orientations of NV centers in our data. For the nanodiamond shown in Figure 1 we observe two distinct sets of peaks in the bare state. This scenario arises from a particular orientation where one NV center is aligned with the external field and three are oriented at ~109° with respect to the field. This alignment is imperfect, however, with each of the three non-aligned NV groups experiencing a slightly different field. This manifests in our data, where the linewidth of the outer (fully aligned) peaks in the bare state data is narrower than the inner (~109°) overlapping orientations (8.6 MHz vs 10 MHz). This allows us to estimate the broadening which comes from the misalignment of the overlapping orientations as ~1-2

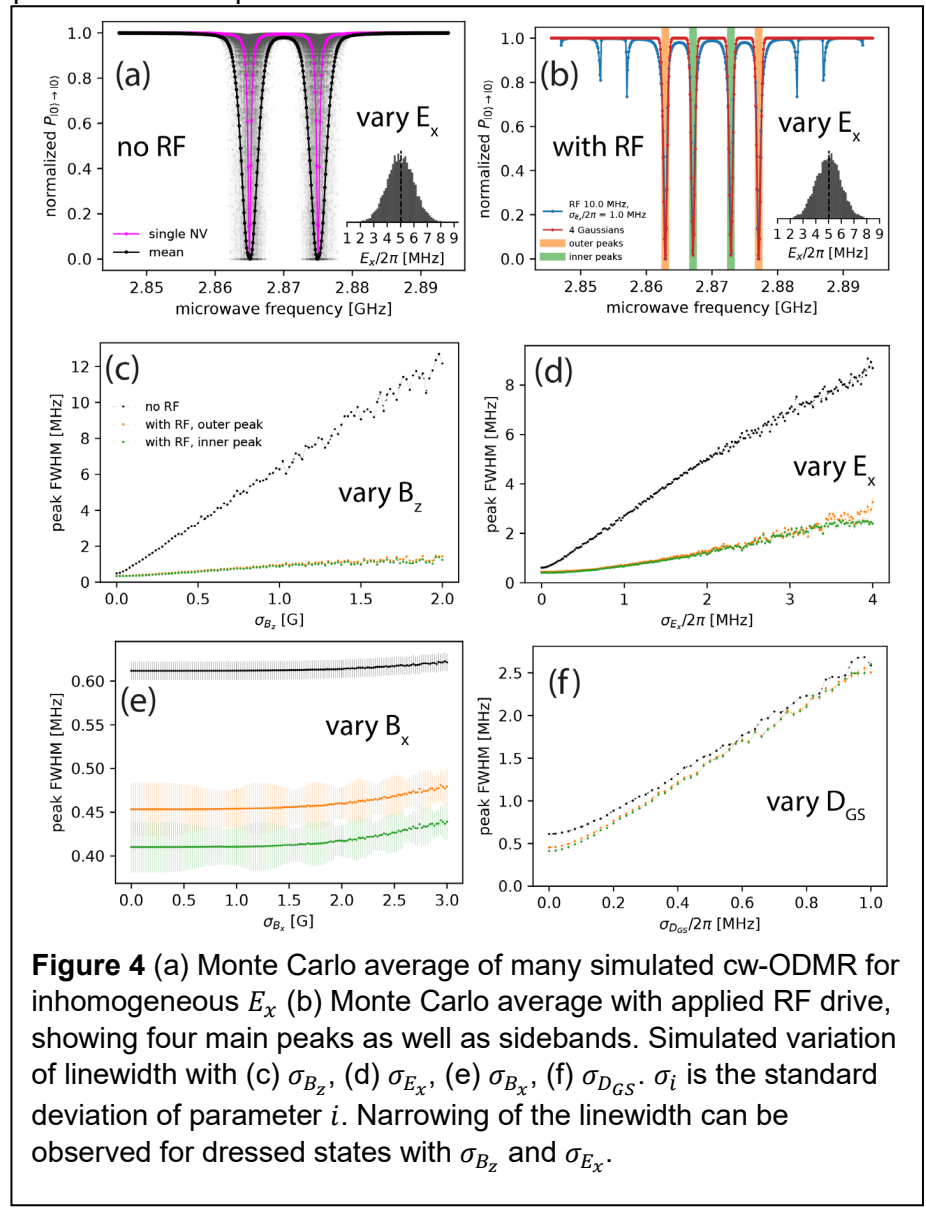

**Figure 4** (a) Monte Carlo average of many simulated cw-ODMR for inhomogeneous $E_x$ (b) Monte Carlo average with applied RF drive, showing four main peaks as well as sidebands. Simulated variation of linewidth with (c) $\sigma_{B_z}$, (d) $\sigma_{E_x}$, (e) $\sigma_{B_x}$, (f) $\sigma_{D_{GS}}$. $\sigma_i$ is the standard deviation of parameter $i$. Narrowing of the linewidth can be observed for dressed states with $\sigma_{B_z}$ and $\sigma_{E_x}$.

MHz. The dressed states will *not* mitigate this source of broadening in the peak. However, orientations which allow each group to be separately addressed would lower the contrast by a factor of three from our observed value, which would likely outweigh any gain in sensitivity from narrower linewidths.

Guided by the simulations of line broadening, we consider the prospects for improving dressed-state temperature sensing through tuning material properties of the nanodiamonds. The sensitivity is governed primarily by four distinct components: the contrast (peak depth), linewidth, photon detection rate, and temperature-dependent frequency shift (see Equation 3). The linewidth of the ODMR transition is one of the major differences in the properties of NV centers in large single crystal diamond substrates, where linewidths of <1MHz can readily be observed [76,77]. We observe that the dressed states can narrow the linewidth by almost a factor of two in many cases (see Table 1), but this is still significantly broader than previous dressed state linewidths in bulk diamond of ~0.7 MHz [25,42]. Our simulations predict that the dressed states should be robust in mitigating the effect of both strain (ubiquitous in nanodiamonds) and static magnetic fields (from e.g., P1 paramagnetic impurities), but suggest $D$-strain as a source of broadening which could be mitigated with improvements in the fabrication process to minimize strain. We note that even if a measurement protocol could be designed to suppress the effect of variations in $D_{GS}$, this would not be desirable for temperature sensing since the variation of $D_{GS}$ with temperature is the mechanism of detection.

The contrast we observe for the dressed states is similar (though not identical) to those in the bare states, as expected from our simulations and previous results in large single crystals. The contrast in both cases is limited fundamentally by photophysical rates of the NV center [47], and practically by other fluorescent background (including that from NV centers which are not resonant with the microwave field). Removing background fluorescence and charge state instabilities is a route to improve the sensitivity of both dressed state and bare NV centers. Contrast and linewidth could also be improved by diamond with NV centers along a single lattice direction (i.e. preferential orientation), although this is difficult to do for high concentrations of NV centers [18] and preferential orientation has only been demonstrated in bulk diamond, not in nanodiamonds.

**Conclusions**

We have demonstrated the generation and characterization of dressed states in NV center-containing nanodiamonds and evaluated their potential as temperature sensors. These dressed states can be generated in nanodiamonds over a range of different magnetic field environments, and we find that these states are able to suppress the magnetic field sensitivity of NV centers by a factor of seven, enabling us to improve the specificity of these sensors to temperature changes. We simulate the sensitivity of the dressed states to several environmental and material parameters to explore the origins – and limitations – of the line narrowing we observe when we generate these dressed states. The ability to engineer Hamiltonians tailored to specific sensing challenges makes NV centers – and related quantum sensors – promising solutions to the long-standing challenge of sensing specificity.

**Experimental Methods**

For nanodiamond imaging and ODMR data acquisition we used a home-built confocal fluorescence microscope equipped with a scanning galvanometer system. A 532nm wavelength laser (LaserQuantum) is used for excitation. Emission is detected with an avalanche photodiode (Excelitas SPCM-AQRH-14-FC). The microwave (GHz) and RF (MHz) signals are generated with signal generators (Rohde & Schwartz SML03, Rohde & Schwartz SMIQ 03B), amplified (ZHL-15W-422-S+ for GHz signals, ZX60-100VH+ for RF signals, gain +36dB, max output 30dBm), and combined with a power combiner (ZFRSC-183-S+). The combined RF and microwave signals continued to the stripline antenna fabricated on a silicon chip, which was terminated with a 50-Ohm coaxial RF terminator.

Nanodiamonds (Adamas NDNV140nmHi) were drop cast onto a 5 mm x 5 mm silicon chip patterned with a gold stripline, which was mounted on a temperature-controlled stage. The temperature-controlled stage was heated with two HT15W resistive cartridge heaters, monitored on the witness site with a TH100PT platinum resistance temperature detector, and controlled with a TC200 temperature controller, which had a resolution of 0.1 K. Prior to the experiment, the witness point was calibrated with another TH100PT sensor in the sample position, then removed and the silicon chip put in place and wire-bonded. During the experiment the temperature controller was set to each temperature setpoint and allowed to thermally equilibrate for at least 10 minutes.

The Monte Carlo simulations for examining linewidth broadening mechanisms with and without RF were performed using a numeric implementation of a Floquet expansion method [40,66]. This method takes advantage of the time-periodicity $\hat{H}(t + T) = \hat{H}(t)$ of the Hamiltonian (Eq. 2) to re-cast it as a Fourier expansion of a time-invariant Hamiltonian; this expansion is then truncated to the desired order and the Schrödinger equation solved using exact diagonalization. To simulate an ODMR spectrum, we iterate through microwave frequencies and at each $\omega_{MW}$ we diagonalize and find the eigenstates of the Floquet Hamiltonian. We then find the probability for staying in the $|0\rangle$ state (i.e. the probability of not transitioning to the $|B\rangle$ or $|D\rangle$ states). We used a Floquet order of $N = 4$ as higher orders were numerically close and the linewidth was only measured for the four largest peaks corresponding to the lowest-order transitions (Fig. 4b). All Floquet simulations presented in this work use $\langle P_{0\to 0}\rangle$ to simulate ODMR contrast.

**Supplemental Information for Selective temperature sensing in nanodiamonds using dressed states**

**Observation of dressed states in other nanodiamonds.**

Figure S1 shows additional data demonstrating the generation of dressed states in other nanodiamonds, highlighting the ubiquity of this approach. Each nanodiamond has a random orientation with respect to the lab frame, resulting in different relative drive strengths for the microwave and radiofrequency AC fields. Overlapping peaks obscure the fine structure around the avoided crossing in several cases, but the response of the nanodiamond to the applied RF field is clear in all cases.

Figure S2 shows the magnetic field dependence of these nanodiamonds in the absence of a dressing drive. Variations in peaks splitting and peak intensities arise from the different projections of the static B-field and microwave B-field respectively.

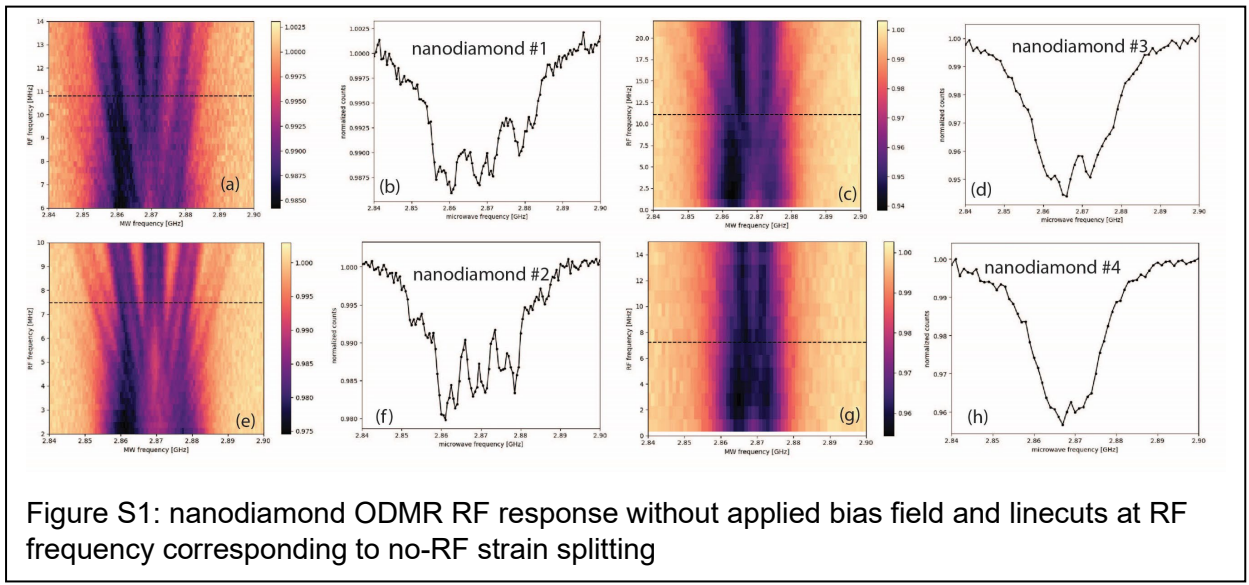

Figure S1: nanodiamond ODMR RF response without applied bias field and linecuts at RF frequency corresponding to no-RF strain splitting

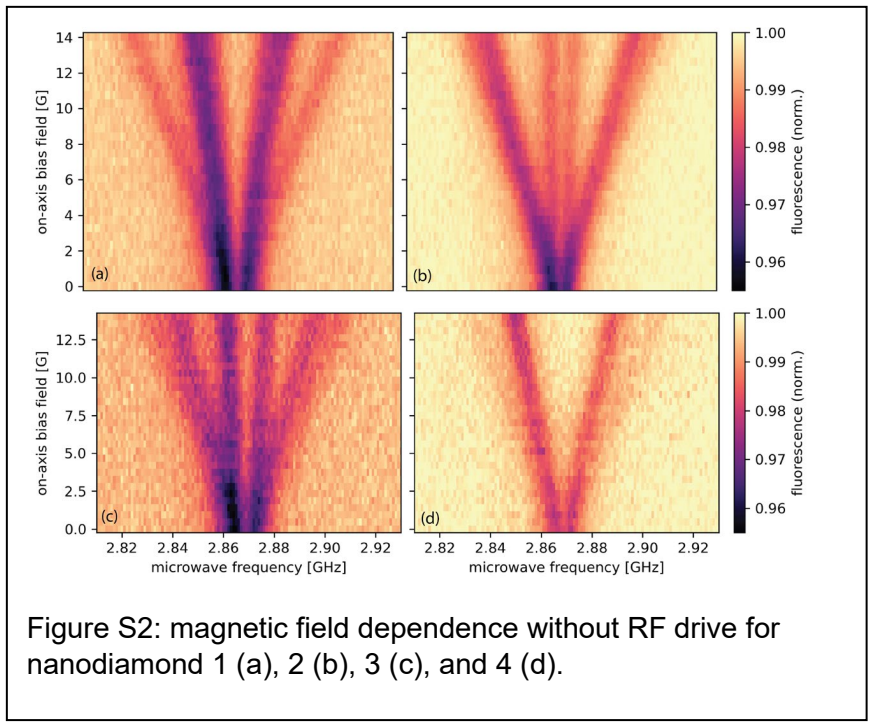

Figure S2: magnetic field dependence without RF drive for nanodiamond 1 (a), 2 (b), 3 (c), and 4 (d).

**Magnetic Field Dependence of Dressed States**

The model we use to describe the data in Figure 3d in the main text uses parameters determined from other experiments rather than allowing these to vary as fit parameters. Table S1 summarizes the values and sources of each term in the equation used to describe the data.

$$\omega_{MW} = D_{GS} \pm \frac{1}{2}\omega_{RF} \pm \sqrt{\left(\sqrt{(\gamma_e B \cos\theta)^2 + E_x^2} - \frac{1}{2}\omega_{RF}\right)^2 + \left(\frac{\Omega_{RF}}{2}\right)^2}$$

| Symbol | Description | Value | Source |
|---|---|---|---|
| $D_{GS}$ | Zero-field splitting | 2.8696 GHz. | Fit to ODMR magnetic field dependence |
| $\omega_{RF}$ | Angular frequency of RF drive | $2\pi$ 25 MHz. | Set by microwave source |
| $\gamma_e$ | Gyromagnetic ratio of electron | $2\pi$ 2.80 MHz/G. | Constant |
| $\cos\theta$ | Directional cosine along NV center for lattice direction | -1/3 | Fit to ODMR magnetic field dependence for inner peaks |
| $E_x$ | Off-axis strain amplitude. | $2\pi$ 5.08 MHz | Fit to ODMR magnetic field dependence |
| $\Omega_{RF}$ | Rabi frequency in the z-direction for RF drive. | $2\pi$ 5.9 MHz | Peak splitting for D2a and D2b at 25 MHz and 9.8 G |

Table S1: fit parameters for model of peak shift with bias field (see Figure 2d).

Table 2 in the main text summarizes the best temperature sensitivities for each nanodiamond in the absence of an applied magnetic bias field but does not give the value of each parameter used to compute temperature sensitivity via equation 3. Table S2 supplies these values as well as the RF power and RF frequency used in the dressing drive.

| ND | RF? | best $\eta_T$ [K/√Hz] | dD/dT [kHz/K] | δ(dD/dT) [kHz/K] | FWHM [MHz] | δFWHM [MHz] | Peak height [A.U.] | max Mc/s | MW power [dBm] | RF power [dBm] | RF freq [MHz] |
|---|---|---|---|---|---|---|---|---|---|---|---|
| 1 | no | 2.7 | -87.2 | 6.5 | 9.1 | 0.3 | -0.06 | 1.4 | -30 | n/a | n/a |
| 1 | yes | 1.8 | -95.3 | 3.3 | 12.3 | 0.4 | -0.09 | 0.7 | -20 | -14 | 25 |
| 2 | no | 4.4 | -93.5 | 10.1 | 15.7 | 0.5 | -0.05 | 2.0 | -30 | n/a | n/a |
| 2 | yes | 2.7 | -92.6 | 14.4 | 16.0 | 1.2 | -0.08 | 0.8 | -20 | -14 | 15 |
| 3 | no | 4.1 | -88.1 | 5.3 | 9.4 | 0.4 | -0.05 | 1.0 | -30 | n/a | n/a |
| 3 | yes | 2.3 | -116.2 | 24.4 | 11.0 | 1.2 | -0.06 | 0.6 | -20 | -14 | 20 |
| 4 | no | 5.5 | -87.2 | 5.0 | 9.7 | 0.5 | -0.04 | 1.0 | -30 | n/a | n/a |
| 4 | yes | 2.8 | -102.1 | 15.1 | 11.6 | 1.2 | -0.07 | 0.3 | -20 | -14 | 15 |

Table S2: CW-ODMR fit parameters used to calculate temperature sensitivity for best values reported in Table 2. Uncertainties in FWHM are derived from statistical output of the Lorentzian peak fits and uncertainties in $\partial D/\partial T$ are derived from linear fits of average peak position from 25°C to 50°C (see Figure 2c). MW power and RF power are given as values before amplification.